\documentclass{article}

\usepackage{graphicx}

\title{The International Linear Collider:\\Prospects and Possible Timelines}

\author{Joachim Mnich\footnote{Talk presented at the International Workshop on Future Linear Colliders (LCWS2018), Arlington, Texas, 22-26 October 2018. C18-10-22.}\\
        Deutsches Elektronen-Synchrotron~(DESY)\\Notkestr.~85, D-22607~Hamburg, Germany\\
        E-mail: {\em joachim.mnich@desy.de}}

\date{}

\begin{document}

\maketitle

\begin{abstract}
The case for the International Linear Collider and the prospects for its realisation as well as possible timelines are discussed.
\end{abstract}

\section{The International Linear Collider}

The field of particle physics since long is facing a number of fundamental challenges, among them the solution of the puzzle of dark matter, the source of the matter-antimatter asymmetry observed in the universe, the structure of the vacuum or the explanation of electroweak symmetry breaking, and the hierarchy problem are among the most prominent ones. 
Some of these challenges can be addressed at future high-energy colliders that have the potential to dicovery "physics beyond the Standard Model".
Of particular importance are detailed studies of the 125 GeV Higgs boson and on precision measurements of the top quark, as well as direct searches for new particles and phenomena at the highest energy scales. 

Any future machine designed to address these problems must fulfil the following requirements: The required precision calls for a high-luminosity lepton collider; a centre-of-mass energy suffcient for top quark pair production and for studies of the top-Yukawa coupling and the Higgs self-coupling (requiring at least 500 GeV) is essential. This criterion excludes circular accelerators from further considerations, since they are limited in energy reach by synchrotron radiation losses. The possibility to extend a future machine to even higher energies is highly desirable, as is the possibility to have polarised lepton beams. Furthermore, the overall cost of any future project as well as its overall power consumption are important factors to be considered. Note that the cost for a linear collider scales linearly with energy, while for a circular collider, rather a quadratic dependence of cost on energy is to be expected. Also, the timescale on which a future collider might be realised is of importance. It is worth noticing that a linear collider also offers the opportunity to replace existing accelerator technology by novel technologies like drive beams or plasma wakefield accelerators, which promise significantly higher gradients.

\begin{figure}
\begin{center}
\includegraphics[width=1.0\textwidth]{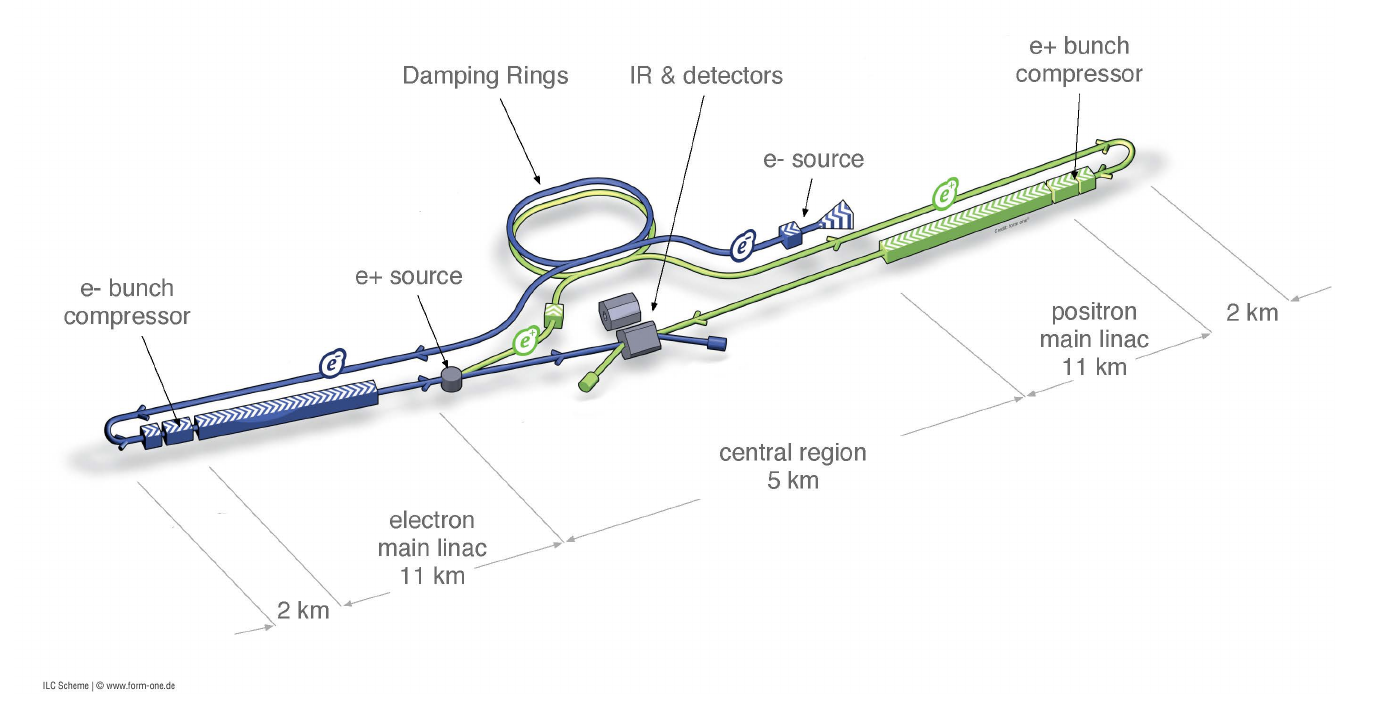}
\caption{ILC baseline design for 500 GeV centre-of-mass energy~\cite{ilctdr}.}
\label{ILC:baseline}
\end{center}
\end{figure}

\begin{figure}
\begin{center}
\includegraphics[width=.9\textwidth]{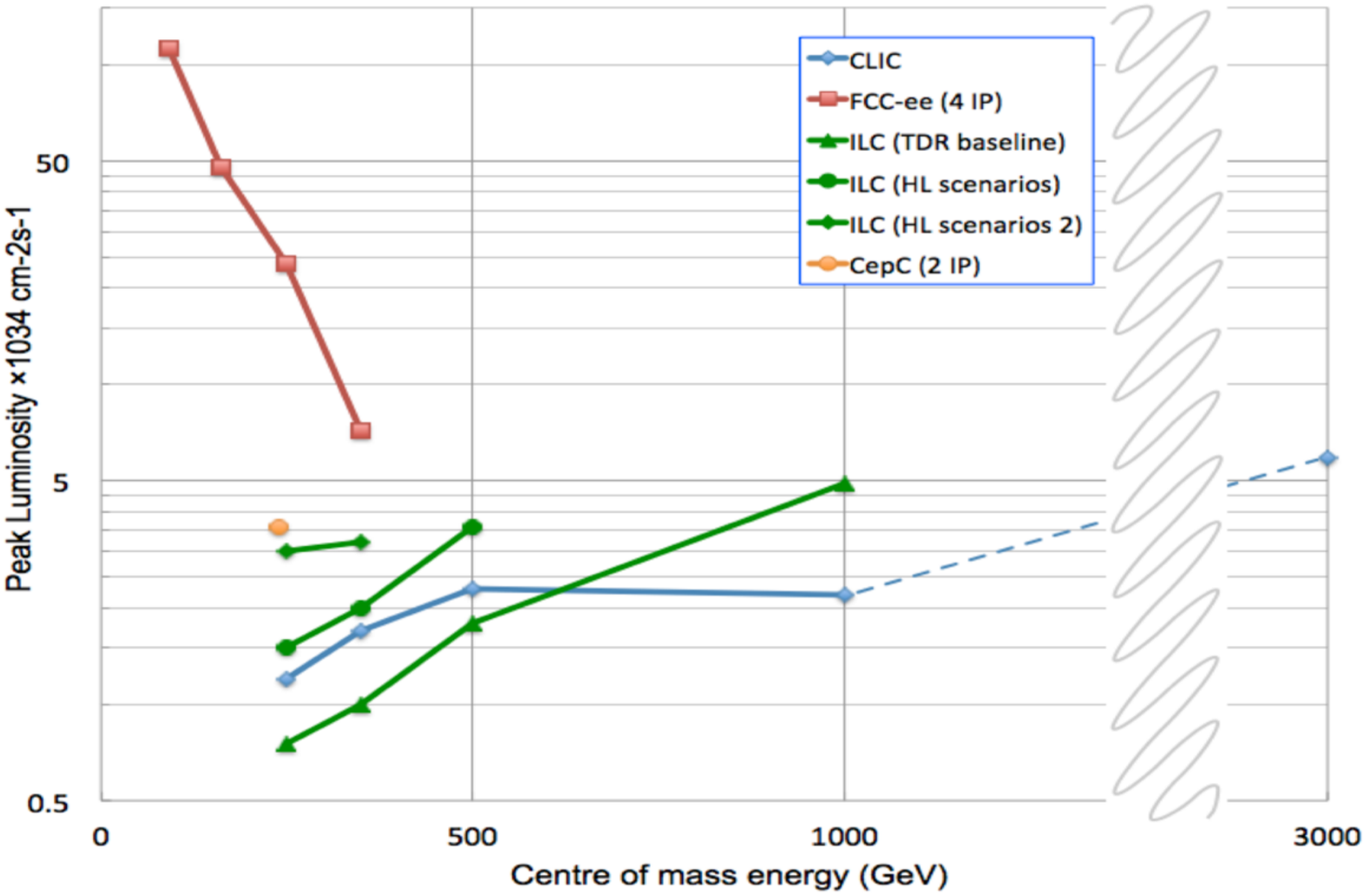}
\caption{Luminosities and centre-of-mass energies of different proposed electron-positron colliders. Source: N.~Walker/DESY.}
\label{ILC:lumi}
\end{center}
\end{figure}

The International Linear Collider (ILC) was specifically conceived to significantly advance particle physics as a whole. Designed as a linear electron-positron collider with a centre-of-mass energy of between 250 and 1000 GeV and a luminosity, at 500 GeV, of at least $1.8 \times 10^{34} cm^{-2}s^{-1}$, it addresses many of the above-mentioned challenges. The accelerator design uses superconducting radio-frequency niobium cavities operating at 1.3 GHz (cooled by 2 K helium). For the 500 GeV machine, about 16000 cavities with Q factors of better than $10^{10}$  and average gradients above 35 MV/m are foreseen to be organised into 1800 cryo-modules, over a length of about 30 km (8000 cavities on about 20 km for the 250 GeV version). The technical design report (TDR) of the ILC was published in 2013~\cite{ilctdr}, containing also an estimate of 163 MW power consumption for the 500 GeV machine. Figures~\ref{ILC:baseline} and~\ref{ILC:lumi}, respectively, show the baseline layout of the collider and a comparison of the achievable luminosity versus centre-of-mass energies for different proposed electron-positron colliders.

The ILC offers a rich physics programme at all foreseen energies --- 250, 500, and 1000 GeV --- based on Higgs precision physics, top-quark physics and searches for physics beyond the standard model. In particular the centennial discovery of the Higgs boson in 2012 opened a new window towards  physics beyond the known and can well be used as a handle for this purpose. 
So far, at the LHC ``new physics'' has been unsatisfactorily absent, and so it seems that --- besides high energy that allow direct access to higher scales --- precision as available at a lepton collider becomes key to BSM physics. In this repect, the ILC can achieve what is out of reach at the LHC, e.g.\ measurements of Higgs-boson couplings to the percent level. The ILC is indeed a Higgs factory at all centre-of-mass energies. Even at 250 GeV, the very clean HZ final state allows access to many Higgs-boson properties. 
In short, the ILC, especially when taken together with the results from the LHC and the HL-LHC, provides excellent prospects on physics beyond the standard model. 

\begin{figure}[ht!]
\begin{center}
\includegraphics[width=.71\textwidth]{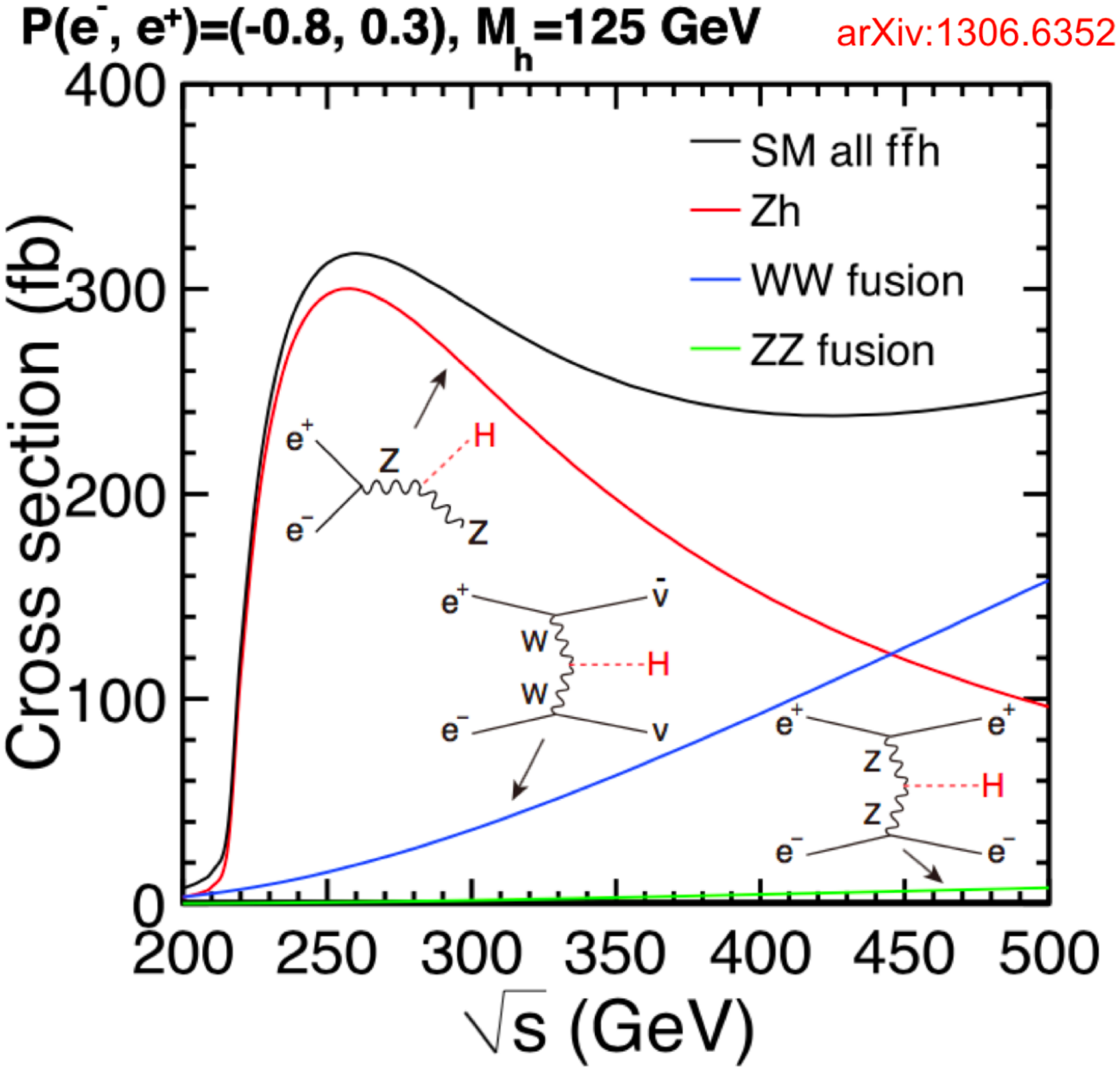}
\caption{Higgs boson production cross sections at the ILC~\cite{ilchiggspaper}. }
\label{ILC:higgs}
\end{center}
\end{figure}

The ILC is technically mature, and at the European XFEL (see below), the feasibility of industrialised component production has been demonstrated. Still, even if there are no technological showstoppers for the ILC, a number of technical challenges still have to be solved: In order to achieve the desired luminosity of  $1.8 \times 10^{34} cm^{-2}s^{-1}$, extremely small beam sizes have to be realised at the interaction point. This so-called ``nano-beam scheme'' requires beams with about 6 nm width in the vertical direction. Experiments at KEK have demonstrated that this goal is in reach~\cite{nanobeam}.
Further technological challenges are the positron source and the beam dump. 
Also on the side of future detectors for the ILC, things at a mature stage: There exist two well-established concepts --- ILD and SiD --- with elaborate solutions for all components. 

Therefore, all in all, we are convinced that we could build the ILC basically now, and justification for this confidence can be found at DESY in Hamburg: the European XFEL (see below).

\section{The European XFEL}

\begin{figure}
\begin{center}
\includegraphics[width=1.0\textwidth]{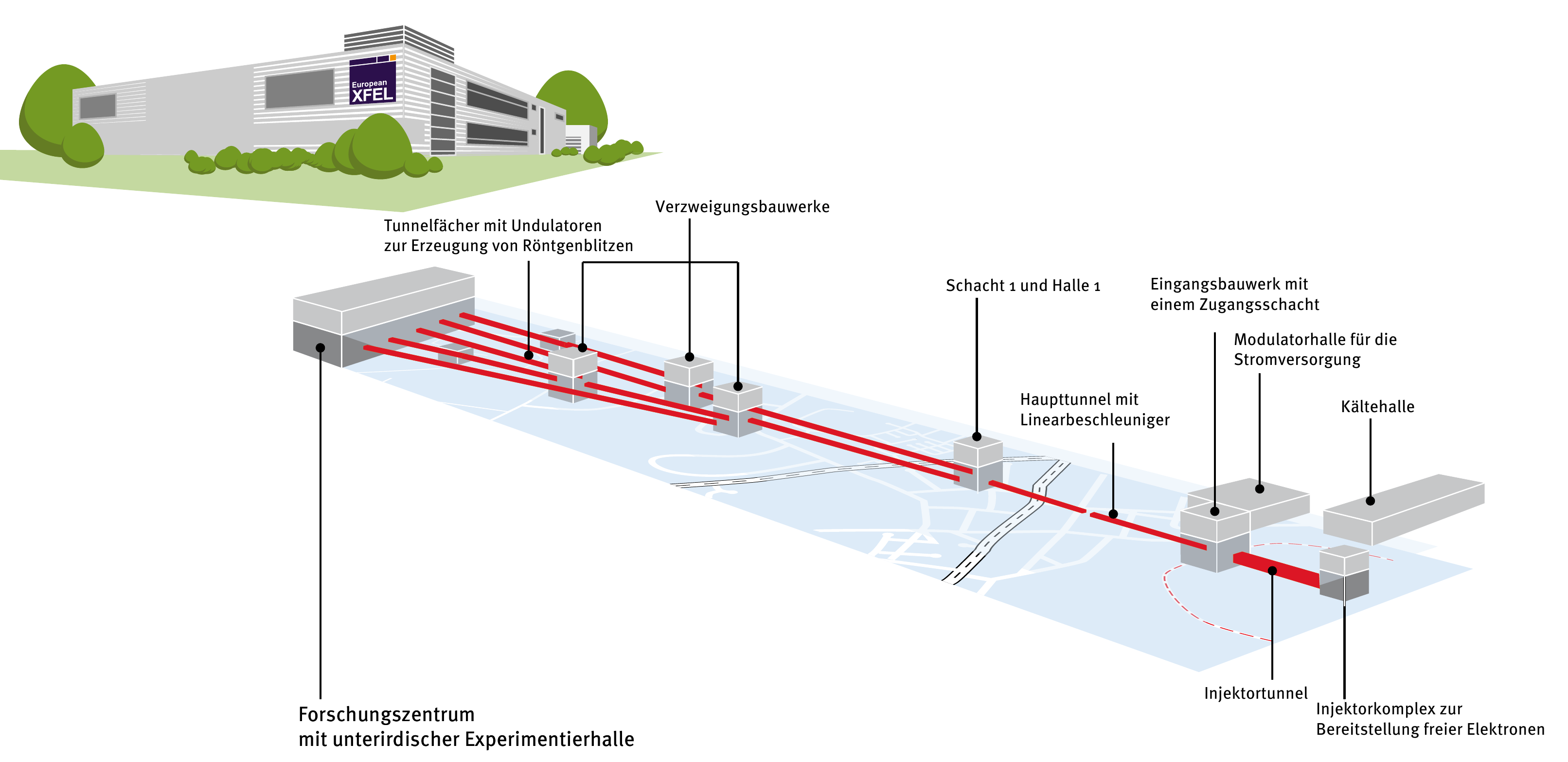}
\caption{The European XFEL facility in Hamburg. Source: Britta von Heintze/Welt der Physik.}
\label{ILC:xfelpicture}
\end{center}
\end{figure}

The European XFEL~\cite{xfelweb} is a 3.4 km long X-ray laser facility. The main linac of the facility, which employs the superconducting radio-frequency technology developed for the ILC, has a length of 1.7 km and a design energy of 17.5 GeV. The European XFEL can thus be considered a ``prototype'' of the ILC. The machine has been under construction since  early 2009 when the European XFEL GmbH was founded; in 2010 the accelerator consortium was founded, consisting of 16 institutions coordinated by DESY. In 2012 the tunnel was finished and installation of the infrastructure started. In 2016, the accelerator was completed and commissioning started with the cool-down of the accelerator. Routine user operation began in September 2017, after first laser light had been observed in early May 2017. Figure~\ref{ILC:xfelpicture} shows an overview of the machine. 

\begin{figure}[hhh]
\begin{center}
\includegraphics[width=1.\textwidth]{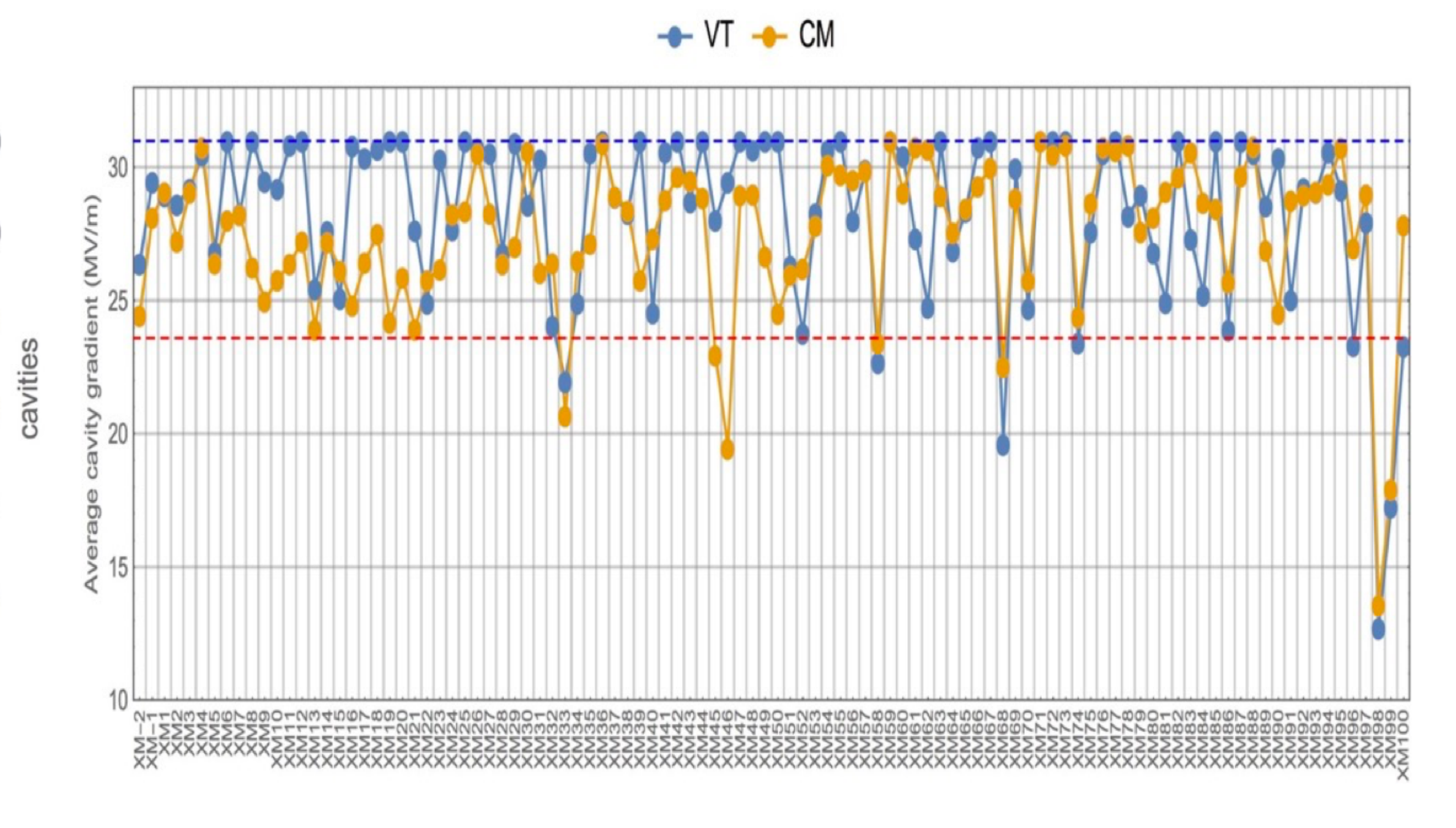}
\caption{High-voltage test results of European XFEL accelerator cavities (``VT'') and cryo-modules (``CM''). Source: N.~Walker/DESY.}
\label{ILC:xfelcavities}
\end{center}
\end{figure}

The European XFEL features 100 cryo-modules with altogether 800 niobium cavities, with an average accelerating gradient of 23.6 MV/m.  Industrialisation of cavity production and the sharing of the work on the cryomodules among European partners were crucial ingredients of the successful construction process. The experience gained is very promising, raising confidence that for the ILC the contemplated sharing of production among world regions and with many partners is the right model. Figure~\ref{ILC:xfelcavities} shows high-voltage test results of the European XFEL cavities and cryo-modules. It can be seen that European XFEL specifications were achieved; additional test results indicate that extrapolation to ILC results will be possible. 

\section{The ILC --- International Support}

The ILC is by definition an international project. New machines are multi-billion Euro enterprises, so that with all probability, there will always only be one of a kind. International consensus is therefore mandatory.  In order to achieve this consensus, coordinated strategy processes worldwide are conducted, and the last full round of such processes concluded 2013 (update of European strategy for particle physics, US HEPAP P5 recommendations, etc.). These processes take  different flavours in different regions of the world, but overall, the picture of a global coherent strategy with ILC as high priority project emerges. We are just facing the next round of these strategies processes, with the beginning update of the  European strategy expected to be published in May 2020 (see later); the USA is expected to follow two or three years later. 

ICFA --- the International Committee for Future Accelerators~\cite{icfa,royspaper} --- was created in 1976 by the International Union of Pure and Applied Physics (IUPAP).  Its mandate is to promote international collaboration on very high energy accelerators, to regularly organise world-inclusive meetings for the exchange of information on future plans, and to organise workshops for the study of problems related to super high-energy accelerators. ICFA is the recognised body to represent high energy physics on a global scale. Its aim is to facilitate international collaboration in the planning, construction and exploitation of accelerators for high energy physics and related fields.

ICFA has been supporting the ILC --- or more generally the idea of a high-energy linear electron-positron collider --- since the 1990s. In 2002, ICFA created the International Linear Collider Steering Committee (ILCSC) to promote the construction of an electron-positron linear collider through world-wide collaboration. In 2003, the International Technology Recommendation Panel (ITRP) was created, and in 2005 ICFA set up the Global Design Effort (GDE) to produce an ILC design and an estimate of its cost. Then, in 2013, after the publication of the ILC technical design report (TDR) the ILCSC came to an end, and the Linear Collider Board (LCB) was formed by ICFA in order to oversee the Linear Collider Collaboration (LCC) that includes both ILC and the Compact Linear Collider (CLIC) study. 

In November 2017, ICFA made a statement on the 250 GeV version of the ILC, emphasising that the ILC is a ``key science project complementary to the LHC and its upgrade''.~\cite{icfaottawa}. ICFA also welcomed the efforts by LCC on ILC cost reduction that comes with additional R\&D and with a reduction of centre-of-mass energy to 250 GeV, but stressed ``the extendability of the ILC to higher energies'' and noted that ``there is large discovery potential with important additional measurements accessible at energies beyond 250 GeV.'' ICFA therefore strongly encouraged ``Japan to realize the ILC in a timely fashion as a Higgs boson factory with a center-of-mass energy of 250 GeV as an international project''. 

But the ILC is supported not only by ICFA. Strong statements in favour of the ILC are also issued by numerous national HEP communities and sometimes even government officials. In Germany, for example, the ``Committee for Elementary Particle Physics'' as the recognised body representing the HEP community, in its input statement to the European strategy update says: ``An electron-positron collider, upgradeable to a centre-of-mass energy of at least 500 GeV, should be realised, with highest priority, as the next international high-energy collider project. ... We strongly support the Japanese initiative to realise, as an international project in Japan, the ILC as a ``Higgs-Factory'' with an initial centre-of-mass energy of about 250 GeV.'' Also scientists in France have expressed (and since long shown) strong interest for the ILC; they were also seriously involved in the European XFEL and with detector R\&D. Similarly,the Italian INFN in its 2015 white paper ``What next?'' supported participation in studies and R\&D related to future colliders: ``Our community must be part of the planning of the future.'' 

For the USA, J.~Siegrist, Associate Director for HEP at the Department of Energy in May 2018 stated that the US looks forward to a decision this year by Japan to host the ILC as an international project. Similary, US Under-secretary Paul Dabbar is cited in the Kitakami Times of 17 October 2018 as ``I hope for the Japanese government to proactively appraise the ILC project and move it forward.'' 
 
In Europe, the community has started the next update process of the European strategy for particle physics. Community input --- also on the ILC, a potential European contribution to it, and on the ILD and SiD detector concepts (see e.g.\ \cite{ilcglobal}) --- had been collected by December 2018, and the next steps in the process foresee a community ``townhall meeting'' or open symposium in Granada, Spain, in May 2019, the publishing of a ``physics briefing book'' digesting that input by September 2019, and a strategy update drafting session by the European Strategy Group (ESG) in January 2020 in Bad Honnef, Germany. The final strategy update is to be approved and published by CERN Council in May 2020. 

\section{ILC --- Status, Model and Possible Timeline}

In autumn 2012, the Japanese high-energy physics community expressed its interest to host the ILC in Japan. It followed a very busy year 2013:  The ILC TDR~\cite{ilctdr} was published, the newly updated European strategy for particle physics supported the ILC project (as did, shortly after, the US ``Snowmass/P5 strategy"" process), and a site in the Kitakami mountain region in the Iwate prefecture was suggested by the Japanese community as a place for realising the ILC in Japan. The Japanese MEXT ministry started a detailed evaluation process of the ILC project, setting up, in May 2014, an ``ILC Advisory Panel'' and working groups for the detailed evaluation of the physics case, the validation of the technical design report, and for a review of the human resources situation. It soon became clear that the project was considered too expensive; cost reduction became a serious issue. 

\begin{figure}
\begin{center}
\includegraphics[width=.8\textwidth]{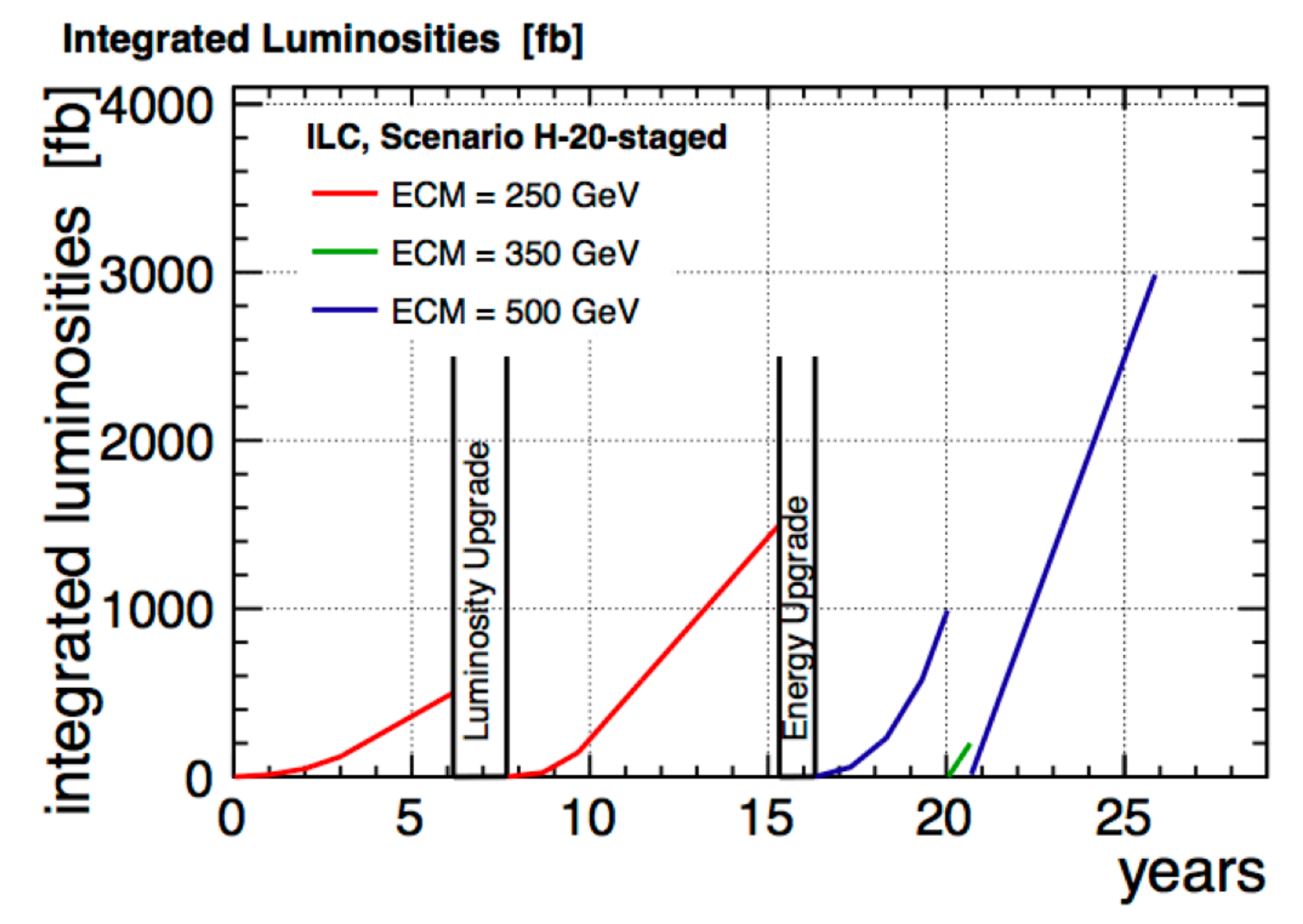}
\caption{Possible ILC running scenario with a starting energy of 250 GeV and subsequent energy upgrades~\cite{ilceft}. }
\label{ILC:scenario}
\end{center}
\end{figure}

The most significant measure towards cost reduction was the staging of the project, i.e.\ the reduction of the initial energy and thus length of the collider from 500 to 250 GeV (and to about 20 km). Although already the 250 GeV machine offers exciting physics prospects, extendability to higher energies was always assumed. Figure~\ref{ILC:scenario} shows a possible running scenario with uns of 250, 350 and 500 GeV that would allow all ILC physics goals to be reached. 

The costs for the 250 GeV machine are supposed to amount to 6.35-7.03 billion USD~\cite{newcost} (excluding detectors, land purchase, preparation phase costs, operation costs). Of this amount, about 1 billion USD are for civil engineering / construction, about 4 billion USD for the accelerator, and the rest for institutional labour in the contributing labs. The detectors are prized at around 1 billion USD in total, and the annual operation costs of the facility at under 400 million USD.  

MEXT, in July 2018, then produced a report on the ILC~\cite{newcost}, on which the Science Council of Japan commented in late 2018~\cite{scjreport}, triggering reactions by the community, the  Japanese High Energy Accelerator Research Laboratory KEK, and the Japanese Federation of Diet Members for the ILC~\cite{reactions}. This Science Council statement together with other deliberations and information will form the basis for the decision to be taken by the Japanese government on the realisation of the ILC. Originally, this decision was expected before the deadline for European strategy update process in December 2018, and it is now expected for early March 2019 when ICFA and the LCB will meet next to discuss future directions. 

In parallel, discussions on possible international support for the ILC in Japan between Japan and other countries (in particular the US, France, and Germany) were conducted on the parliamentary, ministerial and expert level --- it is clear that international consensus and a sharing of the significant cost are mandatory for the success of a large project like the ILC. 

Assuming a positive decision to host the ILC from the Japanese government, and assuming that the ILC is ranked among the high-priority items on the European (to be published in May 2020) and other strategies, international negotiations and the detailed planning of the project might begin. For these preparatory steps, a period of at least four years is foreseen, during which construction will be prepared, the deliverables to the project will be defined, and the ILC international laboratory will be founded. 
After this preparation period,  construction of the accelerator and, in parallel, of the detectors could begin. First physics might then, optimistically, be expected for the mid-2030s --- just in time to replace the HL-LHC as an international flagship of particle physics. 

Potential fields of European interest in contributing to the ILC have been described in the ``European International Linear Collider Preperation Plan''~\cite{eilcpp}, which was prepared in the framework of the E-JADE project~\cite{ejade}, and they were also included in the input to the European strategy update~\cite{europeanactionplan}. These fields are to a large extent defined by capacities and expertise built up during and for the European XFEL project. However, preparations in Europe and elsewhere will start only if Japan sends a sufficiently strong positive signal.

\section{Summary and conclusions}

We have seen strong overall and global support for the ILC --- potentially the next large high-energy collider after the LHC. In realistic scenarios, the ILC could initially be working as a Higgs factory at a reduced centre-of-mas energy of 250 GeV before being upgraded, after some years, to higher energies. 

The project is scientificially exciting --- for many it is the best-suited concept for addressing the pressing questions of particle physics --- and has been shown to be technically mature and feasible within a well-understood budget and timescale. Nevertheless, as in any large project, numerous technological and industrial challenges will have to be mastered on both the accelerator and the detector side. 

At the time of writing this report, the particle physics community is eagerly waiting for a statement from the Japanese government on the realisation of the ILC in Japan under Japanese leadership. This statement is expected latest by early March 2019. Then, high-level inter-governmental discussions on the ILC could start. Still, the realisation of the ILC will require continous enthusiasm and support from scientists and industrial partners, and a global consensus among politicians and funding agencies.

2019 will be a decisive year for the ILC

\section*{Acknowledgement}

I would like to thank Thomas Sch\"orner for help in preparing this document.


\end{document}